\documentclass[conference]{IEEEtran}
\ifCLASSINFOpdf
\else
\fi
\usepackage{graphicx}
\usepackage{soul}
\usepackage[linesnumbered,ruled]{algorithm2e}
\usepackage{subfig}
\usepackage[fleqn]{amsmath}
\usepackage{relsize}
\begin{document}
\relscale{0.9}

\title{An Operating System Level Data Migration Scheme in Hybrid DRAM-NVM Memory Architecture\vspace{-.5cm}}

\author{\IEEEauthorblockN{Reza Salkhordeh and Hossein Asadi}
\IEEEauthorblockA{Data Storage Systems \& Networks (DSN) Lab, Department of Computer Engineering\\
Sharif University of Technology, Tehran, Iran\\
Email: salkhordeh@ce.sharif.edu and asadi@sharif.edu}\vspace{-.7cm}
}

\maketitle
\begin{abstract}
With the emergence of \emph{Non-Volatile Memories} (NVMs) and their shortcomings such as limited endurance and high power consumption in write requests, several studies have suggested hybrid memory architecture employing both \emph{Dynamic Random Access Memory} (DRAM) and NVM in a memory system.
By conducting a comprehensive experiments, we have observed that such studies lack to consider very important aspects of hybrid memories including the effect of: a) data migrations on performance, b) data migrations on power, and c) the granularity of data migration.
This paper presents an efficient data migration scheme at the Operating System level in a hybrid DRAM-NVM memory architecture. 
In the proposed scheme, two \emph{Least Recently Used} (LRU) queues, one for DRAM section and one for NVM section, are used for the sake of data migration.
With careful characterization of the workloads obtained from PARSEC benchmark suite, the proposed scheme
prevents unnecessary migrations and only allows migrations which benefits the system in terms of power and performance.
The experimental results show that the proposed scheme can reduce the power consumption up to 79\% compared to DRAM-only memory and up to 48\% compared to the state-of-the art techniques.
\end{abstract}

\IEEEpeerreviewmaketitle

\section{Introduction}
\label{sec:introduction}
\vspace{-0.15cm}
In the past decades, computer designers have steadily used \emph{Dynamic Random Access Memory} (DRAM) as the main memory due to its prominent features such as high performance and low cost per GB. 
Despite of the performance and cost efficiency of DRAM, it still suffers from frequent recharge requirement and low scalability.
Recharging DRAM cells every few milliseconds imposes significant power, no matter how many accesses are dispatched to the main memory.
The power usage of DRAM is more pronounced when system is mostly idle.
In addition, the low scalability of DRAM limits the maximum main memory size that can be used in a computer system \cite{itrs}.

To alleviate the limitations of DRAM, \emph{Non-Volatile Memories} (NVMs) have been emerged in the recent studies offering zero leakage current to preserve data and less scalability issue as compared to DRAM.
Among various NVMs offered in the past years, \emph{Phase-Change Memory} (PCM), \emph{Spin-Transfer Torque} (STT-RAM),
and \emph{resistive RAM} (PRAM) are recognized as the most promising NVMs to be employed in the main memory \cite{raey}.
Despite prominent features of NVMs, they have serious shortcomings such as high dynamic write power and long write latency (similar to solid-state drives \cite{tiering}) which prohibit them to entirely replace the DRAM technology.
NVMs have asymmetric characteristics for read and write requests.
In most emerging NVMs, write requests require more time for completion and therefore, their performance will be lower in write-dominant workloads.
From power perspective, write requests are more power consumptive than read requests.
In addition, NVMs have very limited write cycles compared to DRAM despite of several efforts to increase their lifetime \cite{sadegh,7155523}.

Due to shortcomings of DRAM, several studies have attempted to employ NVMs in the main memory of computer systems.
A few of these studies explore possibility of entirely replacing DRAM with NVMs \cite{10.1109/ISPASS.2013.6557176,Lee:2009:APC:1555754.1555758}.
A recent study shows that NVMs cannot reach the performance and power consumption of DRAM in the near future \cite{10.1109/ISPASS.2013.6557176}.
Other studies investigate using a hybrid memory composed of both DRAM and NVMs and possible effects on the \emph{Operating System} (OS)
\cite{clockdwf,Dhiman:2009:PHP:1629911.1630086,
Qureshi:2009:SHP:1555754.1555760}.
Hybrid memories try to use characteristics of DRAM and NVM in order to improve performance or power consumption as compared to a DRAM-based main memory.
Clock-DWF \cite{clockdwf} is one of the most recent studies in this field that uses two clock algorithms, one for managing DRAM and another for managing NVM.
This technique tries to move data pages between these two memories in order to reduce the power consumption while maintaining almost the same performance level.
Clock-DWF outperforms previous work such as CLOCK-PRO and CAR which makes it the most optimal technique in the literature.
The simulated results of Clock-DWF over hybrid DRAM-NVM memory lacks considering the effect of the migrations between DRAM and NVM memories. In addition, the effect of moving data pages
between the main memory and the secondary storage has been neglected.
There are also several studies that employ hybrid memory architecture in on-chip memory \cite{Sampaio:2014:EAA:2691365.2691395}, whose discussion is beyond the scope of this work.

This paper presents a data migration scheme in a hybrid memory architecture employing both DRAM and NVM in the main memory.
The main aim of the proposed scheme is reducing the number of non-beneficial data migrations between DRAM and NVM memories to improve both performance and power efficiency.
To this end, we use two \emph{Least Recently Used} (LRU) queues (one for DRAM and one for NVM) and optimize the LRU queue for NVM to prevent non-beneficial migrations to DRAM.
The optimizations in the LRU queue are minimal and therefore the proposed scheme will have almost the same hit ratio as an unmodified LRU.
Contrary to Clock-DWF that each write hit will result in moving the page to the DRAM main memory, in the proposed scheme every hit in the NVM LRU will be treated similar to the LRU algorithm with one difference.
If a page stays in the top pages of LRU for more than a threshold accesses, it will be considered hot and will be moved to DRAM.
Since the cost of moving a data page between two memories is high, using this threshold will prevent non-beneficial migrations that are very likely to occur in previous studies such as Clock-DWF.

Both the proposed scheme and previous studies have been simulated using a framework developed similar to Linux memory management layer.
The performance and power characteristics are extracted from the same source as previous studies.
We also used PARSEC to run the experiments \cite{parsec}.
Since the multi-level caches in CPU affect the distribution of accesses dispatched to the main memory, in this paper we used COTSon full-system simulator \cite{cotson} which is able to simulate a multi-core system with many cache levels.
The experimental results show that the proposed scheme can reduce the power consumption up to 48\% (14\% on average), improve performance up to 70\% (48\% on average), and improve endurance up to 93\% (64\% on average) compared to previous studies.
As compared to a DRAM-based main memory, the power consumption is reduced up to 79\% (43\% on average). 

The rest of the paper is organized as follows.
Section \ref{sec:workloadchar} presents our model for evaluating the performance and power consumption in hybrid memories.
The motivation of this work is discussed in Section \ref{sec:prevwork}.
The proposed data migration scheme is presented in Section \ref{sec:proposed}.
Experimental results are reported  in Section \ref{sec:experiment}.
Finally, Section \ref{sec:conclusion} concludes the paper.

\vspace{-0.2cm}
\section{Performance and Power Models in a Hybrid Memory}
\label{sec:workloadchar}
\vspace{-0.15cm}
This section presents a model for performance and power consumption of  hybrid memories.
The proposed model tries to consider all aspects of computer systems which influence the performance and/or the power consumption.
In addition to the traditional moving pages in case of a miss or evicting a data page, hybrid memories have migrations between two memories.
The migration between two memories depends on the architecture of the hybrid memory.
For the sake of generality, we consider separate memory modules for DRAM and NVM that communicate through \emph{Direct Memory Access} (DMA).
If both memory types can be assembled in one module, the migrations can be done more effectively.
The integrated memory, however, requires hardware modification which is out of scope of this paper.
In the following sections, the performance and power models will be presented.

\begin{table}[t]
\caption{Parameters Description}
\label{tbl:paramdesc}
\scriptsize
\centering
\begin{tabular}{|c|c|}
   \hline
\textbf {Parameter}  & \textbf{Description} \\ \hline   
$P_{Hit_{DRAM}}$ & DRAM Memory Hit Probability \\ \hline
$P_{Hit_{NVM}}$ & NVM Memory Hit Probability \\ \hline
$P_{R_{DRAM}}$ & DRAM Read Access Probability \\ \hline
$P_{R_{NVM}}$ & NVM Read Access Probability \\ \hline
$P_{W_{DRAM}}$ & DRAM Write Access Probability \\ \hline
$P_{W_{NVM}}$ & NVM Write Access Probability \\ \hline
$P_{Miss}$ & Main Memory Miss Probability \\ \hline
$P_{Mig_{D}}$ & Probability of NVM to DRAM Migration\\ \hline
$P_{Mig_{N}}$ & Probability of DRAM to NVM Migration  \\ \hline
$P_{DiskToD}$ & Probability of Moving Page to DRAM due to Page Faults\\ \hline
$P_{DiskToN}$ & Probability of Moving Page to NVM due to Page Faults\\ \hline
$T_{R_{DRAM}}$ & DRAM Memory Read Latency $(s)$ \\ \hline
$T_{R_{NVM}}$ & NVM Memory Read Latency $(s)$ \\ \hline
$T_{W_{DRAM}}$ & DRAM Memory Write Latency $(s)$ \\ \hline
$T_{W_{NVM}}$ & NVM Memory Write Latency $(s)$ \\ \hline
$T_{Disk}$ & Disk Access Latency $(s)$ \\ \hline
$Po_{R_{DRAM}}$ & DRAM Read Dynamic Power $(\eta j)$ \\ \hline
$Po_{W_{DRAM}}$ & DRAM Write Dynamic Power $(\eta j)$ \\ \hline
$Po_{R_{NVM}}$ & NVM  Read Dynamic Power $(\eta j)$ \\ \hline
$Po_{W_{DRAM}}$ & NVM Write Dynamic Power $(\eta j)$ \\ \hline
$PageFactor$ & \# of accesses to memory to write a data page \\ \hline
$AvgStaticPower$ & Prorated Static Power Over All Requests \\ \hline
$StperPage$ & Static Power Consumption of a Page $(\eta j/s)$ \\ \hline
$AccessperPage$ & Average Number of Accesses to Each Page $(1/s)$ \\ \hline
\end{tabular}
\vspace{-0.5cm}
\end{table}

\subsection{Performance Model}
\label{sec:perfmodel}
\vspace{-0.15cm}
The performance model depends on the delay of DRAM and NVM, granularity of eviction, and the delay of migration between memories.
For measuring performance, we use \emph{Average Memory Access Time} (AMAT).
The overhead of migrations will be prorated between all accesses to the memory.
Equation \ref{eq:perf} shows the formula for AMAT.
The description of the parameters is available in Table \ref{tbl:paramdesc}.
In this equation, the first two terms calculate AMAT for all hit accesses in either DRAM or NVM.
The third term considers the page faults.
Since transferring a data page from a disk to the memory will be done with DMA, the delay of
writing data blocks to memory will be overlaid with reading the next data block from the disk.
Therefore, OS only sees the disk delay and in this term we only consider the disk delay.

\begin{figure}[h]
\scriptsize
\vspace{-0.5cm}
\begin{flalign}
\label{eq:perf}
\hspace{-2cm} AMAT &= \nonumber \\
& P_{Hit_{DRAM}} * (P_{R_{DRAM}}*T_{R_{DRAM}}+ P_{W_{DRAM}} * T_{W_{DRAM}})\nonumber \\
  &+ P_{Hit_{NVM}}*  (P_{R_{NVM}}  * T_{R_{NVM}} + P_{W_{NVM}} * T_{W_{NVM}}) \nonumber \\
 &+ P_{Miss} * T_{Disk} \nonumber \\
 &+ P_{Mig_{D}} *PageFactor * (T_{R_{NVM}} +  T_{W_{DRAM}})\nonumber \\
    &+ P_{Mig_{N}}*  PageFactor * (T_{R_{DRAM}} + T_{W_{NVM}}) 
\end{flalign}
\vspace{-0.5cm}
\end{figure}

The last two terms calculate the migration cost between two memories.
Upon occurring a migration, a data page will be read from a memory and will be written to the other memory.
Since the granularity of data pages is quite larger than the actual accesses to memory (typically 4 up to 16B), we use $PageFactor$ that is a coefficient
which converts moving of a data page into the required number of accesses to memory.
The granularity of the moves between disk and memory modules and between two memories is a data page which is typically 4KB or 8KB.
In this paper, we assume 4KB data pages.
Moving a data page from disk to either of memories might result in a migration between two memories.
It depends on the employed algorithm for managing hybrid memory.
The proposed performance model takes into account this type of migrations.

\subsection{Power Model}
\label{sec:powermodel}
\vspace{-0.15cm}
The proposed power model tries to consider every aspect of the hybrid memories in order to provide more accurate and more realistic estimations for the power consumption of computer systems. 
While static power consumption is consumed regardless of the number of arrived requests to the memory, dynamic power is consumed per request sent to the memory.
Our power model considers the migration between two memories and moving pages from disk to either of the memory modules as well as static and dynamic power for servicing requests.

The dynamic power consumption is calculated per access to the memory.
This will result in independency of the power model from application runtime and the memory size.
Therefore, we introduce \emph{Average Power Per Request} (APPR) as a metric for measuring the power as shown in Equation \ref{eq:powerdynamic}.
Similar to the performance model, first two terms calculate the power for all hit accesses to the memories.
The third and fourth terms consider the write power for moving a data page from disk to a memory module.
The last two terms take into account the power effect of the migrations between two memories.

\begin{figure}[h]
\scriptsize
\vspace{-0.5cm}
\begin{align}
\label{eq:powerdynamic}
 \hspace{-0.5cm}APPR & = \nonumber \\
 & P_{Hit_{DRAM}} * (P_{R_{DRAM}}* Po_{R_{DRAM}}+  P_{W_{DRAM}} * Po_{W_{DRAM}}) \nonumber \\
& + P_{Hit_{NVM}}* (P_{R_{NVM}}*  Po_{R_{NVM}}+ P_{W_{NVM}} * Po_{W_{NVM}}) \nonumber \\
& +  P_{Miss} * P_{DiskToD} *PageFactor  * P_{W_{DRAM}}\nonumber \\
& + P_{Miss} *  P_{DiskToN}* PageFactor * P_{W_{NVM}}\nonumber \\
& + P_{Mig_{D}} *  PageFactor *(Po_{R_{NVM}}+ Po_{W_{DRAM}}) \nonumber \\
& +P_{Mig_{N}}* PageFactor *(Po_{R_{DRAM}} + Po_{W_{NVM}}) 
\end{align}
\vspace{-0.5cm}
\end{figure}

Since static power consumption is independent from requests, we introduce a new parameter called
$AvgStaticPower$ which prorates the static power consumption between all requests arrived to the memory in a given time interval.
The reason behind prorating the static power over all requests is that from the OS perspective, the main memory consumes power (including both static and dynamic) for servicing the requests and both of the sources of the power consumption should be considered as the cost of servicing the requests.
For a specific workload, $AvgStaticPower$ is calculated according to Equation \ref{eq:powerstatic}.

\begin{figure}[h]
\scriptsize
\vspace{-0.5cm}
\begin{equation}
\label{eq:powerstatic}
AvgStaticPower_{Page} = \frac{StperPage}{AccessperPage} 
\end{equation}
\vspace{-0.5cm}
\end{figure}

Here, $AvgStaticPower$ can be combined with the dynamic power to form an APPR that models all power aspects of hybrid memories.
It is worthy to mention that the dynamic power consumption is still independent from memory size and workload.
As expected, static power per request is still dependent on memory size and request service rate.

\vspace{-0.2cm}
\section{Motivation}
\label{sec:prevwork}
\vspace{-0.15cm}
Designing hybrid memories and employing both DRAM and NVM memories is discussed in many of previous work.
A group of previous studies tried to use DRAM as a caching layer for NVM memory \cite{Qureshi:2009:SHP:1555754.1555760,6799122,7059057}.
Similar to the other caching techniques, if the locality of the requests drops below a threshold, the performance of the cache will be decreased.
In addition, the algorithms employed in the DRAM cache can be moved into the \emph{Last Level Cache} (LLC) of CPU in order to evict mostly read-dominant data pages \cite{cachereplacementpcm}.

Another group of previous studies, similar to our proposed scheme, use DRAM and NVM at the same level in the memory hierarchy \cite{clockdwf,  Dhiman:2009:PHP:1629911.1630086,Ramos:2011:PPH:1995896.1995911,6855546}.
Many of the these studies require hardware modifications in memory module controllers \cite{Ramos:2011:PPH:1995896.1995911,Dhiman:2009:PHP:1629911.1630086}.
There are also very few software-driven techniques that try to use the existing interfaces between OS and memory modules \cite{6855546,clockdwf}.
%


CLOCK-DWF \cite{clockdwf}, which is one of the most effective techniques in the previous work, is a very similar study to this paper and it outperforms many of the previous studies such as CLOCK-PRO \cite{clock-pro}. 
Hence, we will have an in-depth analysis about its performance and power.
CLOCK-DWF uses two clock algorithms one for each of the memory modules.
Upon occurrence of a page fault, if the request causing the page fault is write, the page will be moved to the DRAM and otherwise it will be moved to the NVM.
The modifications in the clock algorithm enables CLOCK-DWF to find popular and write-dominant data pages and move them to the DRAM memory.
If a write request arrives for a data page residing in the NVM memory, the data page will be moved to DRAM.
Migrating pages between two memories require many accesses to both memories.
But such effect is not considered in CLOCK-DWF which will result in inaccuracy of their model.
In the reminder of this section, we will analyze CLOCK-DWF with respect to the proposed performance and power models.

\begin{figure}
\centering
\includegraphics[scale=0.45]{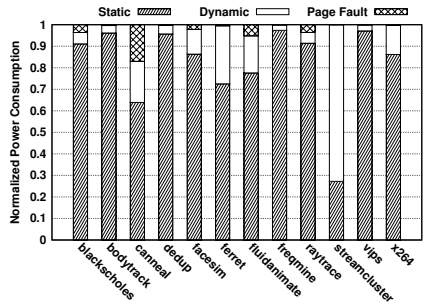}
\caption{DRAM Power Breakdown}
\label{fig:drampower}
\vspace{-0.5cm}
\end{figure}

Before examining CLOCK-DWF, we will calculate the maximum power saving that can be achieved by reducing the static power consumption.
The proposed power model can be used for modeling homogeneous memories.
Hence, the single DRAM main memory is characterized by the proposed model.
Considering a DRAM-only main memory with LRU  algorithm as the eviction policy, Fig. \ref{fig:drampower} shows the composition of the power consumption sources for various workloads.
Since static power consumption contributes for 60-80\% of the total power consumption of DRAM main memory, reducing the static power consumption will have significant effect on the overall systems power consumption.
As shown in Fig. \ref{fig:drampower}, the \emph{streamcluster} benchmark does not behave similar to the other workloads.
According to Table \ref{tbl:workloads}, this workload has a large burst of accesses and a small memory footprint which will result in higher dynamic power consumption.
Workloads with a high hit ratio in LLC of CPU will have higher static power consumption per request.
This is due to less requests will reach the main memory and power consumption will be prorated over fewer number of requests.

CLOCK-DWF maintains two clock algorithms for DRAM and NVM.
The clock algorithm in the NVM is the traditional clock algorithm with one difference.
If a write access arrives for a data page in NVM, the corresponding data page will be moved to the DRAM.
Therefore, no write access will be responded by NVM.
The main aim of this method is to reduce the number of writes in NVM.
Although this prevents any writes from reaching NVM, each write access for a data page in NVM will result in a data page migration between
two memories.
Clock algorithm for DRAM, however, is different and tries to keep write-dominant data pages in the DRAM memory and evicts the mostly read-dominant
data pages.
This is motivated by the fact that the read-only pages will have better performance-power trade-off compared to write requests in NVM.
Upon occurrence of a page fault, if the request is read, the corresponding data page will be moved to NVM and if it is a write, the data page will be moved
to DRAM.

\begin{figure*}[!t]
\centering
\subfloat[]{\includegraphics[width=.3\textwidth]{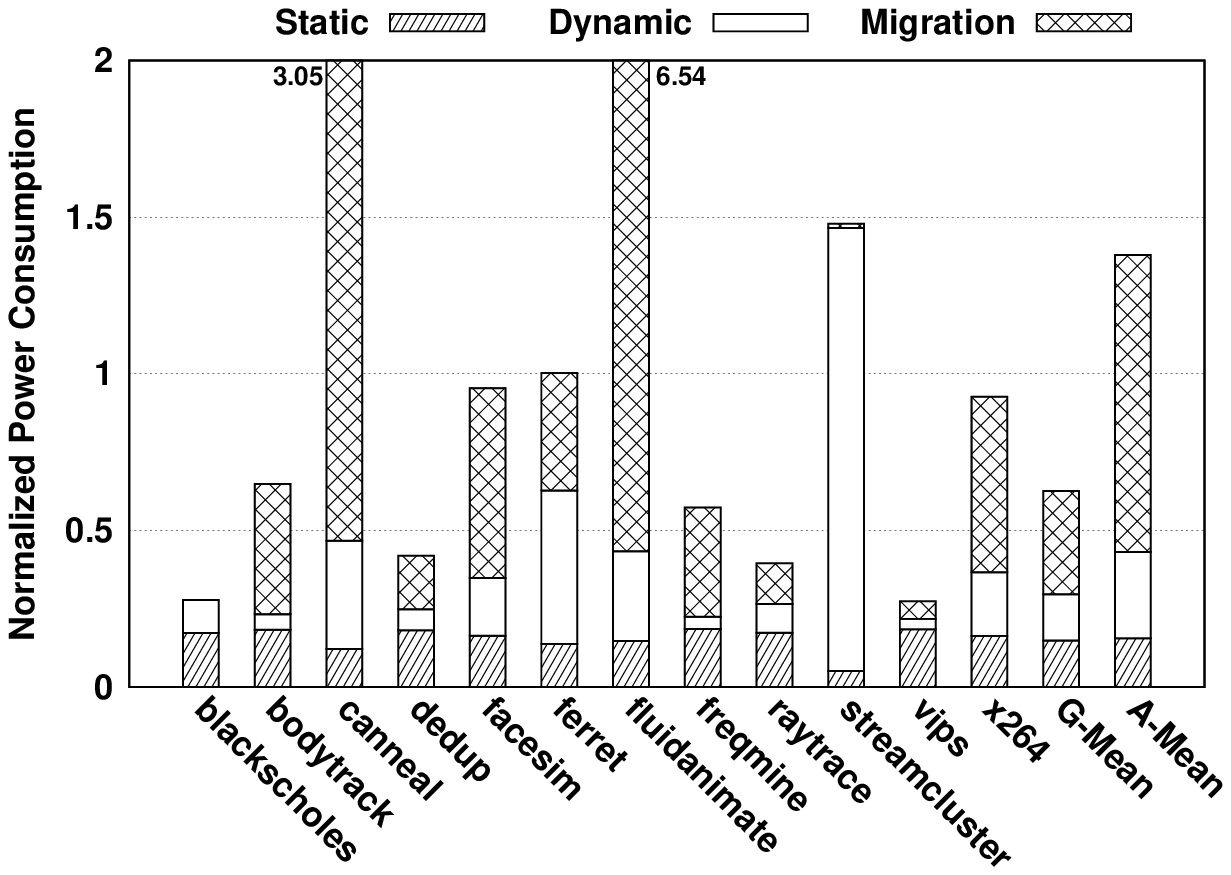}%
\label{fig:clockpower}}
\hfil
\subfloat[]{\includegraphics[width=.3\textwidth]{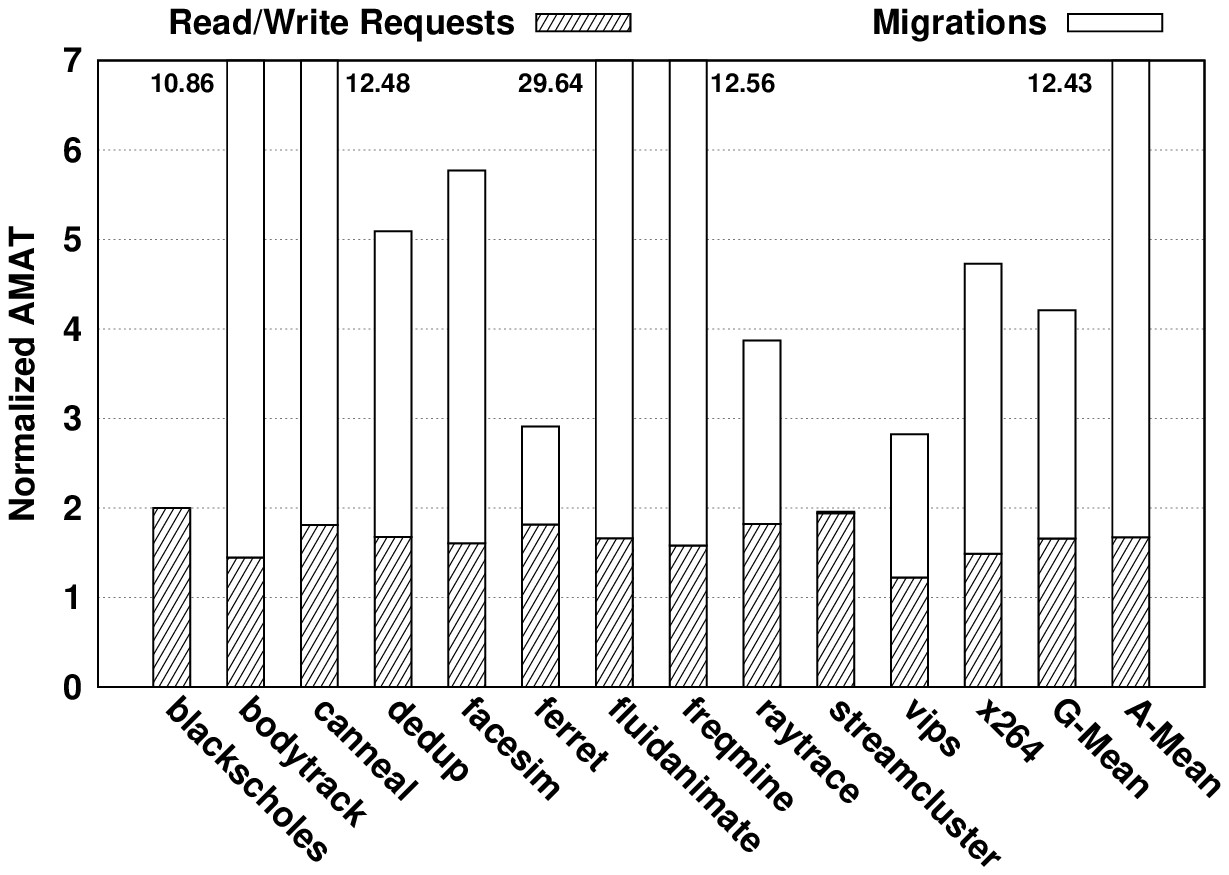}%
\label{fig:clockperf}}
\hfil
\subfloat[]{\includegraphics[width=.3\textwidth]{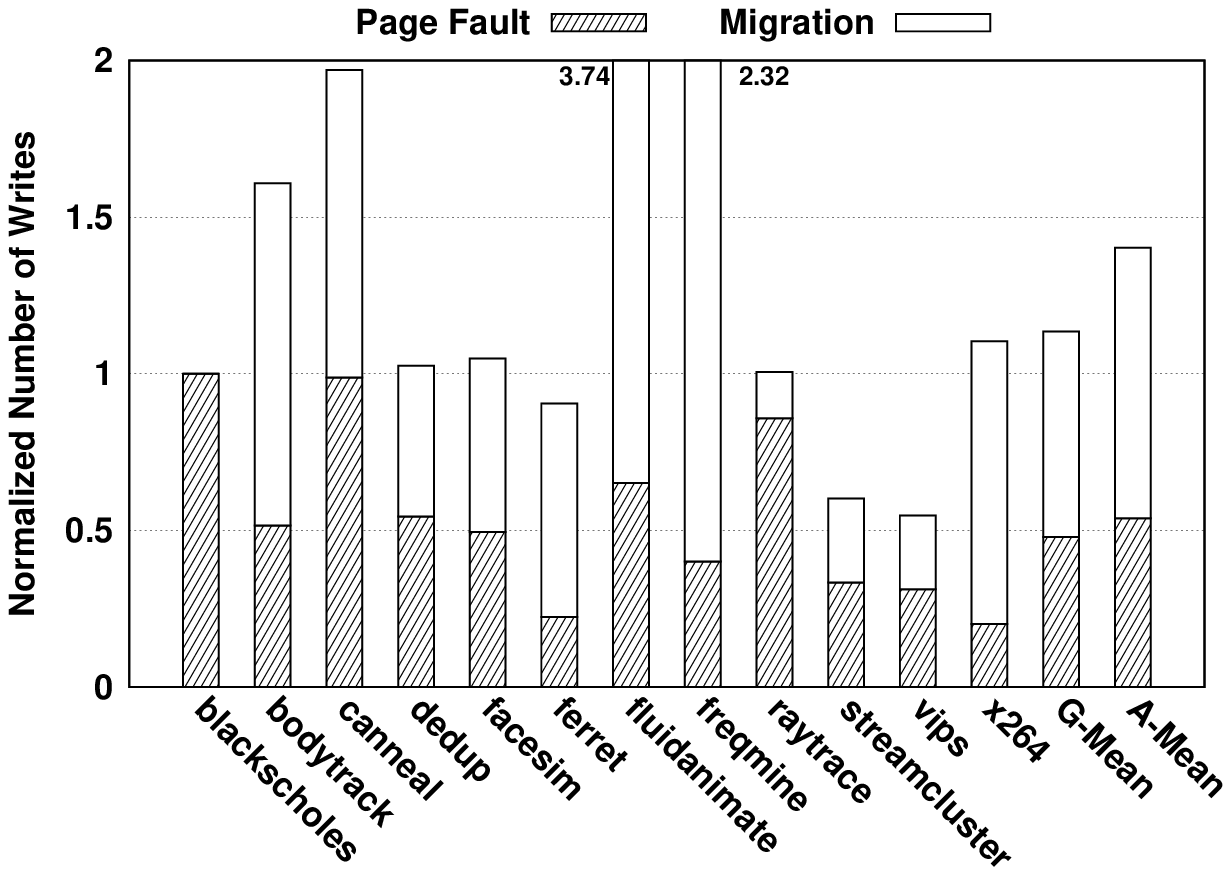}%
\label{fig:clockwrite}}
\caption{a) CLOCK-DWF Power Breakdown Normalized to DRAM Power Consumption b) Normalized AMAT of CLOCK-DWF Compared to DRAM-Only Memory c) Number of Writes in CLOCK-DWF Normalized to NVM-Only Memory}
\label{fig:clock}
\vspace{-0.5cm}
\end{figure*}


\subsection{Power Analysis}
\vspace{-0.15cm}
Fig. \ref{fig:clockpower} depicts the normalized power consumption of CLOCK-DWF compared to power consumption of a DRAM-only memory.
In all workloads, the static power consumption is reduced by 80\% which shows the effectiveness of hybrid
memories to reduce the static power consumption.
Although CLOCK-DWF can decrease the power consumption in many workloads, there are workloads in which CLOCK-DWF fails to improve power consumption and has worse power efficiency compared to DRAM-only memory.
The \emph{streamcluster} benchmark is read-dominant and CLOCK-DWF moves the read-only data pages to NVM.
Therefore, DRAM area will be almost idle and NVM will respond most of the requests.
This will cause the dynamic power consumption to be higher than DRAM-only main memory.
The two other benchmarks that have higher power consumption compared to DRAM are \emph{canneal} and \emph{fluidanimate}.
Although these two workloads are read-intensive, the behaviour of the application causes CLOCK-DWF to migrate a data page to NVM and after a short time, it brings the migrated data pages back to DRAM.
It is worthy to note that the \emph{blackscholes} benchmark is a read-only benchmark and the reason its dynamic power consumption is similar to DRAM-only memory is that
when DRAM is empty, the data page will be moved to DRAM regardless of the type of the request.
In many of the workloads examined in this paper, the contribution of the migrations in power consumption is more than 40\%.
This is due to this fact that when DRAM memory is full, each write access for data pages in NVM will trigger a migration from NVM to DRAM and also,
a migration from DRAM to NVM.


\subsection{Performance Analysis}
\vspace{-0.15cm}
In terms of performance, the source of latencies that can be observed by applications are the delay of responding to the request, the delay of migrations,
and the delay of page faults.
Similar to the power analysis, the performance analysis can show how much we have to pay in terms of latency in order to use a hybrid memory.
Fig. \ref{fig:clockperf} shows the contribution of each source of delay on the AMAT.
AMAT is normalized based on AMAT of a DRAM-only main memory.
The calculated AMAT for requests is very close to the results reported by in the CLOCK-DWF study.
Migrations, however, have not been considered in the CLOCK-DWF study.
Based on the proposed model, the observed delay caused by migrations is considerable and contributes to more than 60\% of the total AMAT.
Therefore, the performance, similar to power, is greatly degraded because of the non-beneficial migrations.
If the hybrid memory algorithm identifies and prevents these migrations, it will reduce the migration cost in terms of performance, power, and endurance.
The beneficial migrations, however, should be allowed to exploit the benefits of hybrid memories.

\subsection{Endurance Analysis}
\vspace{-0.15cm}
As mentioned earlier, CLOCK-DWF does not issue any write requests to the NVM and all writes will be responded in DRAM.
Therefore, the only sources of writes in NVM are migration from DRAM to NVM and moving data pages from disk to NVM in case of a page fault caused
by a read request.
Although the data pages in NVM are read-dominant, each write request for data pages in NVM will result in a high number of physical writes, since the granularity of moving a data page is typically three orders of magnitude larger than the CPU requests.
Fig. \ref{fig:clockwrite} shows the contribution of various sources of writes in NVM.
The number of writes is normalized compared to an NVM-only main memory to see how much CLOCK-DWF can reduce the total number of writes.
In most of the workloads, writes issued for migrations contribute more than 50\% of the total writes in NVM.
This excessive use of migrations makes the overall number of the writes to be even more than an NVM-only main memory.
Hence, the lifetime of NVM will be heavily penalized by using CLOCK-DWF.

\section{Proposed Data Migration Scheme}\label{sec:proposed}
\vspace{-0.15cm}
Non-beneficial migrations are the biggest flaw in CLOCK-DWF and other previous work.
Therefore, in the proposed scheme, we try to identify and prevent this type of migrations.
In addition, the proposed scheme aims to maintain almost the same level of hit ratio as conventional algorithms in order to have comparable performance compared to DRAM-only main memory with LRU algorithm.

The proposed scheme consists of two LRU queues, one queue for DRAM and another queue for NVM.
In order to have a high hit ratio, the algorithms employed in both queues is LRU without any modification.
The proposed scheme manages the migrations between two memories and moves pages from/to disk.
Therefore, both memories work with LRU and the proposed scheme decides when a data page should be migrated to another memory.
Furthermore, upon moving a data page to a memory, it will be treated based on the algorithm of the memory, e.g, moving to the head of the LRU
queue and evicting the last page in the queue.
This is one of the main differences between this work and the previous studies.
In the previous studies, the algorithms for managing pages in memories need to be changed which will result in lower hit ratio.

In order to find the data pages that will improve power consumption and performance upon migration (with respect to the migration cost),
the proposed scheme stores some additional information about data pages such as read and write counters in the NVM LRU queue.
Note that this additional information does not interfere with LRU and it does not need to know about this housekeeping information.
For each data page in the NVM queue, two counters will be stored that count the number of read and write accesses to the corresponding data page
from the time that data page enters the queue.

\begin{figure}
\centering
\includegraphics[scale=0.3]{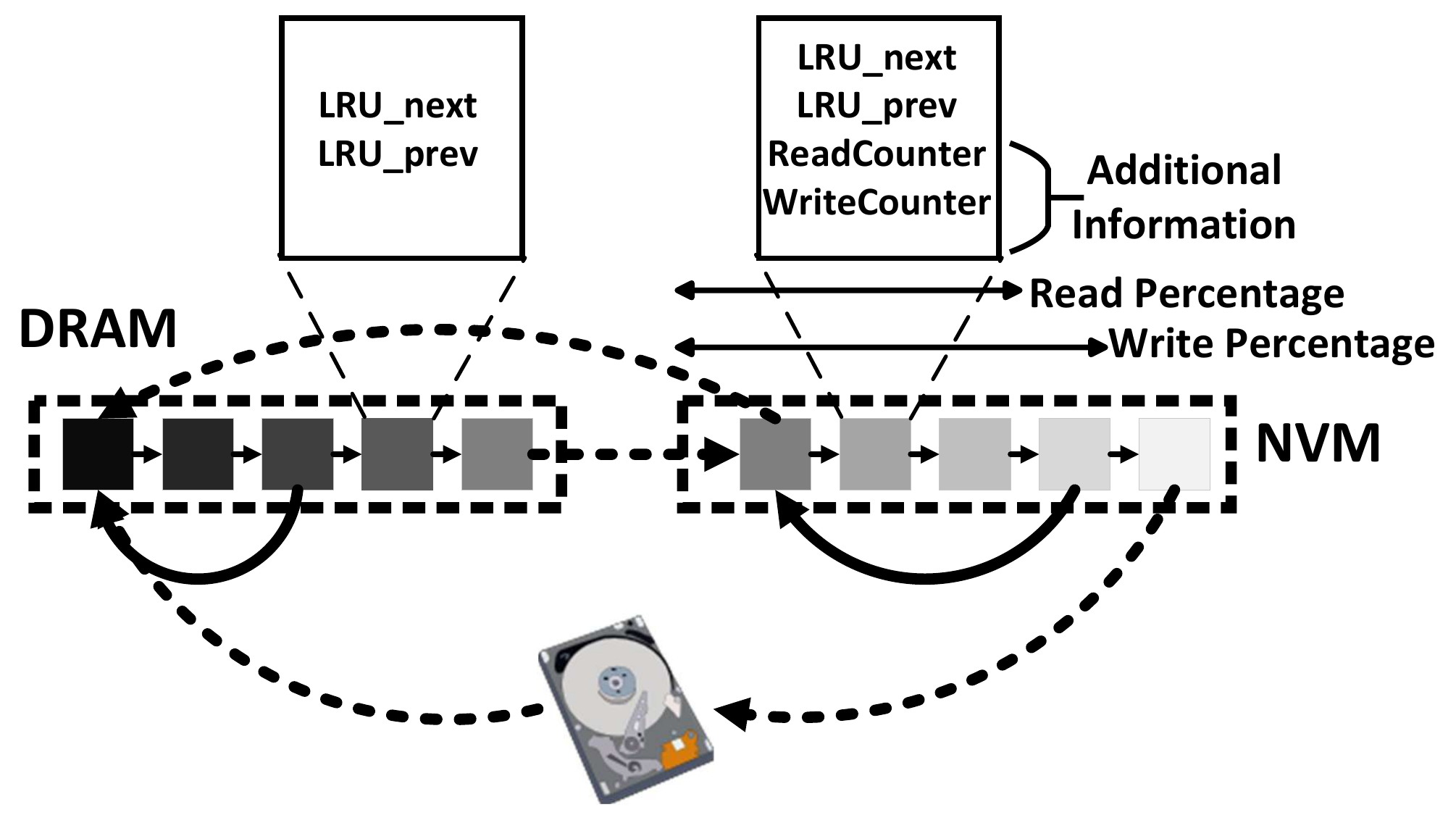}
\caption{Proposed Data Migration Scheme in a Hybrid Memory Architecture}
\label{fig:twolru}
\vspace{-0.5cm}
\end{figure}

Fig. \ref{fig:twolru} shows the architecture of the proposed data migration scheme consisting of two LRU queues.
Dashed lines depict actions performed by the proposed technique and solid line are for traditional LRU management algorithm.
Dark data pages are more frequently accessed and are considered as hot data pages.
Contrary to CLOCK-DWF that places page faults issued by read requests on NVM, the proposed scheme moves all pages from disk to DRAM area.
This is motivated by the fact that moving to either NVM or DRAM will result in a page write in NVM since the DRAM is always full and moving a data page
to DRAM will issue an eviction to NVM.
Therefore, the cost of moving to NVM or DRAM is the same in terms of writes in NVM.
The newly accessed data pages have higher probability of access compared to the older data pages and moving this new page to DRAM will result
in increase in DRAM hit ratio instead of NVM hit ratio.
This will help improving both performance and power efficiency since DRAM is superior in terms of dynamic power and delay.
The overhead of storing the housekeeping information is not considerable and is about 0.04\% for 4KB data pages.
However, keeping the counters for all pages in NVM has a few drawbacks.
First, it requires an ordering scheme in order to identify data pages that are cold but will be accessed once in a long time.
These data pages will reside long enough in NVM to have a high counter values and therefore will be moved to DRAM where they cannot compete
with hot data pages and will return to NVM which makes their migration to DRAM without any benefits.
Second, there is no difference between pages that are frequently accessed and typically reside near the head of the NVM LRU queue for the entire time and
data pages which go back and forth in the queue.

In the proposed scheme, another method has been added to handle both of the above-mentioned issues.
The housekeeping information will be only stored for a few percentage of top positions in the NVM LRU queue.
Once a data page moves to the end of this selected percentage of LRU, the corresponding counter will be reset to zero.
This will handle both ordering scheme and identifying burst data accesses.
Since NVMs have different costs for reads and writes in terms of power and performance, we will treat them differently in the proposed scheme.
Write-dominant data pages should have higher priority over read-dominant data pages for migrating to DRAM since they cost more in NVM.
Therefore, $writeperc$ and $writethreshold$ parameters will be set to higher values than $readperc$ and $readthreshold$.

Algorithm \ref{alg:hybrid} shows the flow of the proposed scheme in case of arriving a request.
Since DRAM contains the most hot data pages, the proposed scheme searches DRAM first and if it is not found, it goes to NVM.
Finding the data page in DRAM will result in a normal LRU housekeeping.
Otherwise, the extra housekeeping information in NVM will be updated based on the request type.
The read and write counters will be stored for $readperc$ and $writeperc$ top data pages in the NVM, respectively.
Therefore, in case of a hit, read and write counters for data pages that are dropped off from the top data pages will be cleared.
Lines \ref{alg:hybrid:nvm5} through \ref{alg:hybrid:nvm4} initialize the counters for the corresponding data page.
If the value of the counter for a data page in NVM exceeds the $read\,threshold$ or $write\,threshold$ (depending on the request type), it will be migrated
to DRAM.
Inserting a new data page into memory and eviction policies are unchanged from LRU and therefore, such details are omitted from the algorithm for the sake of brevity.
The values of $read\,threshold$ and $write\,threshold$ determine how aggressive we plan to prevent the migrations with low probability of being
useful.
It is closely related to the cost of the migration between DRAM and NVM which is related to the performance and power characteristics of the
employed NVM.
\begin{algorithm}[t]
\scriptsize
\SetAlgoNoLine
Search for $request\,address$ in $DRAM\,LRU$\; 
\eIf{$request\,address$ is found in DRAM}{
Update $DRAM\,LRU$\;} 
{Search for $request\,address$ in $NVM\,LRU$\; \label{alg:hybrid:nvm}
\eIf{$request\,address$ is found in NVM}{
Update $NVM\,LRU$\; \label{alg:hybrid:nvm2}
Reset read counter for page in position $readperc$\;\label{alg:hybrid:nvm3}
Reset write counter for page in position $writeperc$\;
\eIf{$request$ is read}{\label{alg:hybrid:nvm5}
\eIf{$request$ is within $readperc$}{
$page\,read\,counter= page\,read\,counter‌ + 1$\;
}{
$page\,read\,counter=1$\;
}
}{
\eIf{$request$ is within $writeperc$}{
$page\,write\,counter= page\,write\,counter + 1$\;}
{$page\,write\,counter=1$\;
}
}\label{alg:hybrid:nvm4}
\If{($request$ is read and $page\,read\,counter > read\,threshold$) or
($request$ is write and $page\,write\,counter > write\,threshold$}{
Migrate page to DRAM\;
}
}{
Issue page fault from Disk to DRAM\;
Migrate from DRAM to NVM if necessary\;
}
}
\caption{Data Migration in a Hybrid Memory}
\label{alg:hybrid}
\end{algorithm}

\section{Experimental Results}
\label{sec:experiment}
\vspace{-0.15cm}
In this section, the experimental setup to extract the traces from workloads and the experimental results for both the proposed method and previous
studies will be presented.

\vspace{-0.1cm}
\subsection{Experimental Setup}
\label{sec:setup}
\vspace{-0.15cm}
The proposed scheme and previous studies are evaluated based on the proposed performance and power models.
For further accuracy of the evaluation, we used COTSon \cite{cotson} which is a full system simulator to obtain memory traces.
The memory traces are extracted from running the actual benchmark programs in a Linux virtual machine inside COTSon and only memory accesses from ROI of the benchmark is considered.
PARSEC-3.0 \cite{parsec} has been selected as the benchmarking suite.
The input of all benchmarks was set to the largest dataset available in order to minimize the effect of starting from cold memory\footnote{swaptions workloads are not included in the results due to compilation issues in our platform.}.

COTSon simulator used a quad-core CPU with two levels of cache and 4GB main memory running an Ubuntu operating system.
Using a quad-core CPU will ensure that there is always enough requests issued to the memory to simulate a production server.
The detailed configuration of the simulated hardware is reported in Table \ref{tbl:cotsonconfig}.
In order to fully understand the effect of different parameters of the workloads on the output of the hybrid memories, the main features of the workloads are presented in Table \ref{tbl:workloads} and will be discussed in the next subsection.
In the experiments, the total memory size is set to 75\% of the total pages and the DRAM size is set to 10\% of the total memory size, similar to previous studies \cite{clockdwf}.
The performance and power characteristics of DRAM and NVM, reported in Table \ref{tbl:memory}, are obtained from the same source as CLOCK-DWF in order to have a fair comparison.

\begin{table}[t]
\caption{COTSon Configuration}
\scriptsize
\label{tbl:cotsonconfig}
\centering
\begin{tabular}{|c|c|}
    \hline
  CPU &  Quad-core with MOESI Protocol \\  \hline
  L1 Data Cache &   32KB WB 4-way set associative with 64B line size  \\ \hline
 L1 Instruction Cache & 32KB WB 4-way set associative with 64B line size  \\     \hline
 Last-Level Cache &   2MB WB 16-way set associative with 64B line size \\    \hline
 Main Memory & 2x 2GB DDR2 \\     \hline
 Secondary Storage & HDD with 5 milliseconds response time  \\     \hline
\end{tabular}
\vspace{-0.25cm}
\end{table}

\begin{table}[t]
\caption{Workload Characterization}
\scriptsize
\label{tbl:workloads}
\centering
\begin{tabular}{|c|c|c|c|}
    \hline
  Workload & Working Set Size (KB) & \# of Read Requests& \# of Write Requests\\  \hline
  Blackscholes & 5,188& 26,242 (100\%) & 0 (0\%) \\ \hline
  Bodytrack & 25,304& 658,606 (62\%) & 403,835 (38\%) \\ \hline
  Canneal &164,768 & 24,432,900 (98\%) & 653,623 (2\%)\\ \hline
  Dedup & 512,460& 17,187,130 (71\%) & 6,998,314 (29\%)\\ \hline
  Facesim &210,368 & 11,730,278 (66\%) & 6,137,519 (34\%) \\ \hline
  Ferret &68,904 & 54,538,546  (89\%) &  7,033,936 (11\%) \\ \hline
  Fluidanimate &266,120 & 9,951,202 (69\%)& 4,492,775 (31\%) \\ \hline
  Freqmine & 156,108& 8,427,181 (69\%)& 3,947,122 (31\%)\\ \hline
  Raytrace & 57,116& 1,807,142 (83\%)&  370,573 (17\%)\\ \hline
  Streamcluster &15,452 & 168,666,464 (99.8\%)& 448,612 (0.2\%)\\ \hline
  Vips & 115,380& 5,802,657  (59\%) &4,117,660  (41\%)\\ \hline
  X264 & 80,232& 14,669,353 (74\%)& 5,220,400 (26\%)\\ \hline
\end{tabular}
\vspace{-0.1cm}
\end{table}

\begin{table}[t]
\caption{Memory Characteristics \cite{clockdwf}}
\scriptsize
\label{tbl:memory}
\centering
\begin{tabular}{|c|c|c|c|}
    \hline
  Memory &   Latency r/w$(\eta s)$  & Power r/w $(\eta j)$ & Static Power $( \frac{j}{GB.second})$ \\  \hline
 DRAM &   50/50 & 3.2/3.2 & 1  \\ \hline
NVM (PCM) & 100/350 & 6.4/32 & 0.1  \\     \hline
\end{tabular}
\vspace{-0.5cm}
\end{table}

\vspace{-0.5cm}
\subsection{Experimental Results}
\begin{figure*}[!t]
\centering
\subfloat[]{\includegraphics[width=.3\textwidth]{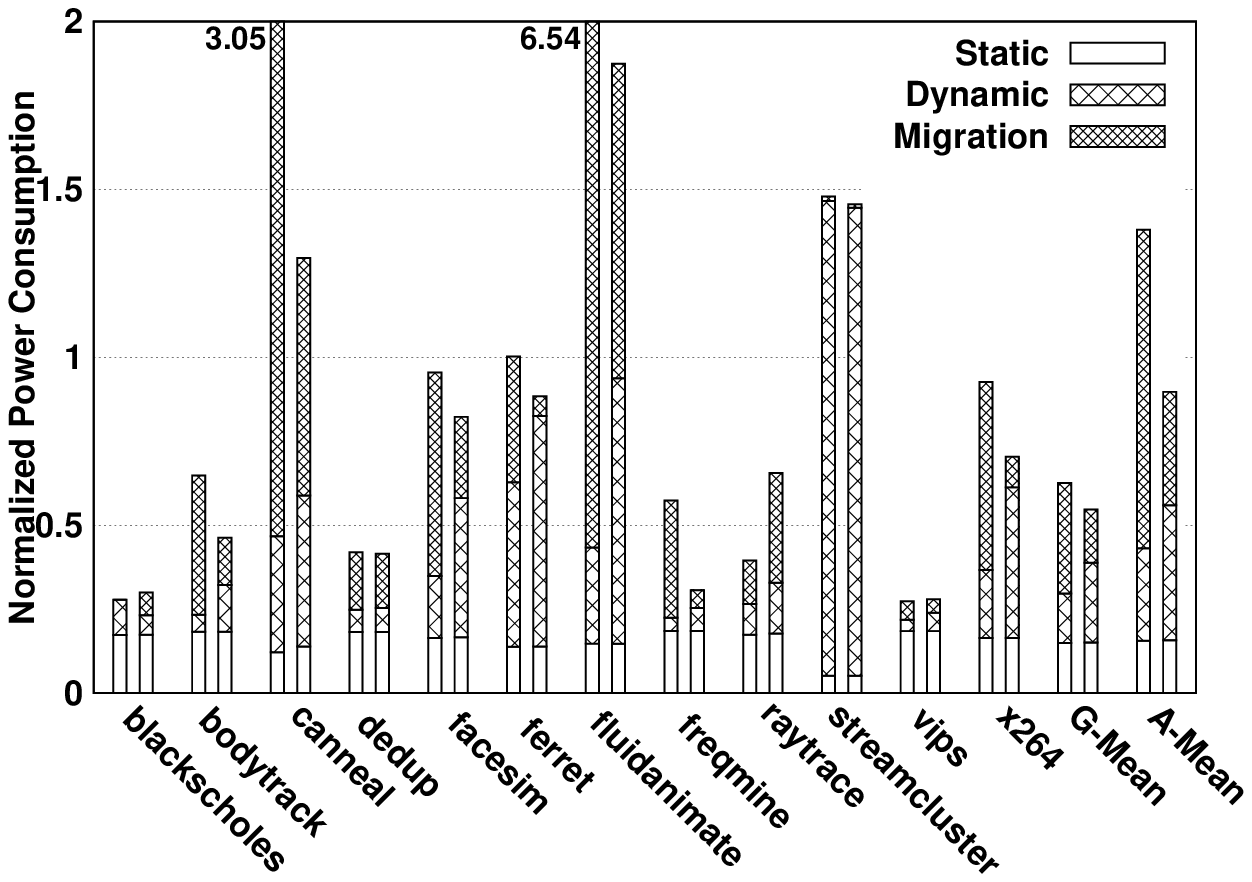}%
\label{fig:proposedpower}}
\hfil
\subfloat[]{\includegraphics[width=.3\textwidth]{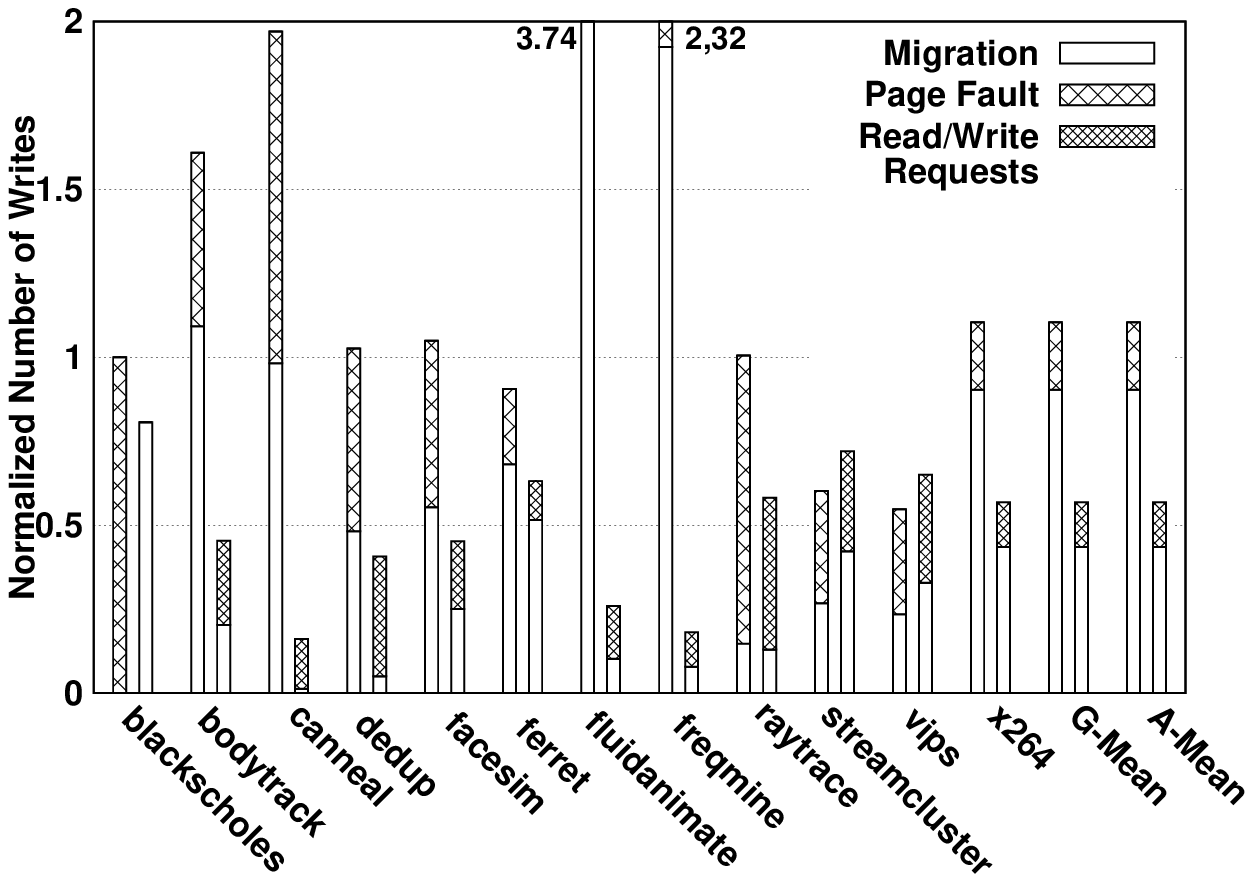}%
\label{fig:proposedwrite}}
\hfil
\subfloat[]{\includegraphics[width=.3\textwidth]{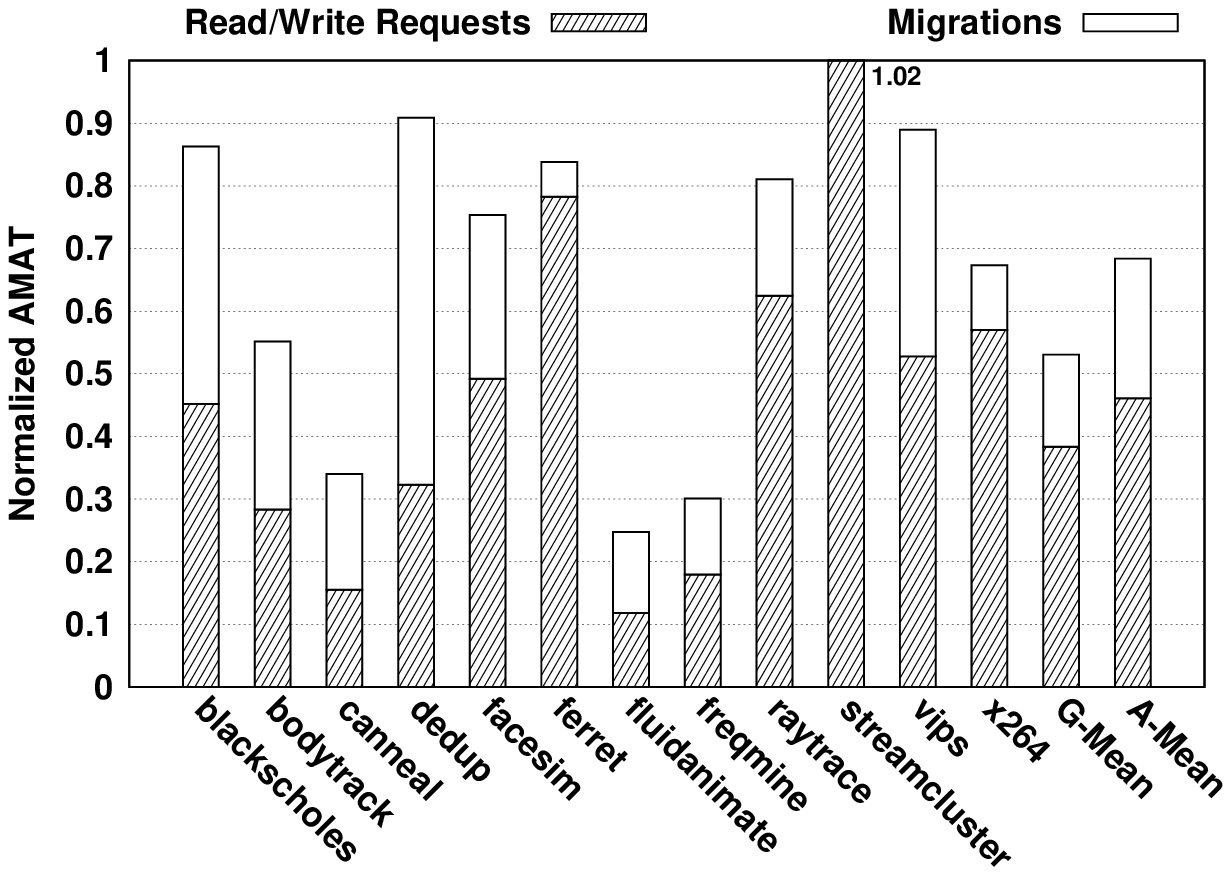}%
\label{fig:proposedperf}}
\caption{a) Power Breakdown of CLOCK-DWF (Left Bar) and the Proposed Scheme (Right Bar) Normalized to DRAM Power Consumption, b) Number of Writes in CLOCK-DWF (Left Bar) and the Proposed Scheme (Right Bar) Normalized to NVM-Only Memory, and c) Normalized AMAT of the Proposed Scheme Compared to CLOCK-DWF }
\label{fig:hitratio}
\vspace{-0.5cm}
\end{figure*}

Fig. \ref{fig:proposedpower} depicts the normalized power consumption of CLOCK-DWF and the proposed scheme compared to a DRAM-only main memory.
For each workload, the left and right bars represent CLOCK-DWF and the proposed scheme, respectively. 
In most of the workloads, the proposed scheme has better power efficiency comapared to CLOCK-DWF with a few exceptions which will be addressed later in this section.
As shown in Fig. \ref{fig:proposedpower}, the power consumption of the proposed scheme is up to 48\% (14\% on average\footnote{Average numbers reported throughout the paper are geometric means.}) less than CLOCK-DWF.
In addition, the proposed scheme can reduce the total power consumption of the main memory up to 79\% (43\% on average) compared to using a DRAM-only main memory.
The static power consumption is the same for both methods since they are evaluated using the same DRAM and NVM size.
The main benefit of the proposed scheme is that the power consumption for migrations is decreased significantly compared to CLOCK-DWF.
The migration cost is decreased up to 80\% by using the proposed scheme.

Among the benchmark programs, \emph{canneal}, \emph{fluidanimate}, and \emph{streamcluster} have unusual characteristics such as small footprint or lack of read-dominant data pages which will increase the dynamic and migration power and makes them not suitable for using hybrid memories.
Contrary to the other workloads, in \emph{raytrace} workload, the migration cost of the proposed scheme is higher than CLOCK-DWF.
Our analysis shows that the optimal values for $readthreshold$ and $writethreshold$ of this workload differs from the other workloads which caused many non-beneficial migrations between two memories.
It is worthy to note that using adaptive threshold prediction can further improve the efficiency of the proposed scheme.
This is part of our ongoing research.

One of the main differences between CLOCK-DWF and the proposed scheme is how they treat write requests attempting to access data pages in NVM.
CLOCK-DWF moves data pages to DRAM while the proposed scheme tries to respond the request from NVM.
Fig. \ref{fig:proposedwrite} shows the normalized number of writes arrived to NVM compared to a NVM-only main memory.
Without considering the migrations, CLOCK-DWF will reduce the number of writes dispatched to NVM.
Considering the migrations, CLOCK-DWF issues more writes to NVM compared to a NVM-only main memory up to 3.7x, which significantly affects the lifetime of NVM.
The proposed scheme, on the other hand, limits the number of migrations between memories and therefore issues less writes to NVM.
The mentioned tradeoff between dispatching requests to NVM and migrating data pages to DRAM affects the contribution of different sources
of writes in NVM.
The proposed scheme favours issuing writes to NVM instead of migrating the whole data page to DRAM while CLOCK-DWF does the opposite.
This change in policy results in significant decrease (up to 93\%) in the number of writes in NVM compared to CLOCK-DWF.
In addition, the proposed scheme can reduce the number of writes in NVM up to 75\% (49\% on average) compared to a NVM-only main memory which will
prolong its lifetime up to 4x.
In \emph{streamcluster} and \emph{vips} benchmark programs, CLOCK-DWF performs slightly better since burst accesses to data pages are near the threshold of being
beneficial migration and the proposed scheme may take a wrong decision on such cases.

From performance perspective, as we concluded in Section \ref{sec:prevwork}, the migrations lead to high delay on the average request
response time in CLOCK-DWF.
Fig. \ref{fig:proposedperf} depicts the normalized AMAT of the proposed scheme compared to CLOCK-DWF.
The proposed scheme successfully limited the number of migrations and the contribution of the migration is less than 50\% in most of the workloads.
Limiting the migrations improves the AMAT of the proposed scheme significantly compared to CLOCK-DWF up to 70\% (48\% on average).
Preventing non-beneficial migrations is not the only reason that the proposed scheme has superior performance compared to CLOCK-DWF.
The policy for selecting the targets for migrations is another reason that the proposed scheme has higher performance than CLOCK-DWF since placing the hot data pages in DRAM will improve AMAT.
In \emph{raytrace} and \emph{vips} benchmarks, CLOCK-DWF has better AMAT since the proposed scheme issues high number of migrations.

\vspace{-0.1cm}
\section{Conclusion}
\label{sec:conclusion}
\vspace{-0.15cm}
NVMs are emerging memory technologies that unlike DRAM, do not have high leakage power and do not depend on the power supply to store data.
NVMs, however, have their own limitations which prevent them from entirely replacing DRAM.
Hybrid memories try to reduce the power consumption of the main memory while maintaining high performance.
Previous studies lack considering all aspects of the hybrid memories and the inaccuracy in their models results in inefficient hybrid memories.
In this paper, we first presented both performance and power models for the hybrid memories. 
Using the proposed models, we identified the shortcomings of previous studies and proposed a novel data migration scheme for hybrid memory.
The proposed scheme consists of two LRU queues with efficient algorithms to manage data migration.
The experimental results show that the proposed scheme can reduce the power consumption up to 79\% compared to DRAM-only memory and up to 
48\% compared to previous studies.

\bibliographystyle{IEEEtran}
\vspace{-0.3cm}
\bibliography{IEEEabrv,Ref}

\end{document}